\documentclass[10pt,letterpaper]{article}
\usepackage{opex3}
\usepackage{color}
\usepackage[colorlinks=true,urlcolor=blue,linkcolor=black,citecolor=black,breaklinks=true,bookmarks=false]{hyperref}
\usepackage{graphicx}
\usepackage{amsmath}
\usepackage{amssymb}
\usepackage{amsthm}
\usepackage{color}
\usepackage{cite}

\newcommand{\be}{\begin{equation}}
\newcommand{\ee}{\end{equation}}
\newcommand{\bea}{\begin{eqnarray}}
\newcommand{\eea}{\end{eqnarray}}

\begin{document}

\title{Coupling to Modes of a \\Near-Confocal Optical Resonator \\Using a Digital Light Modulator}

\author{Alexander T. Papageorge$^{1,3}$, Alicia J. Koll\'{a}r$^{1,3}$, and \\Benjamin L. Lev$^{1,2,3,^*}$}

\address{$^1$Department of Applied Physics, Stanford University, Stanford CA 94305, USA\\
$^2$Department of Physics, Stanford University, Stanford CA 94305, USA\\
$^3$E. L. Ginzton Laboratory, Stanford University, Stanford CA 94305, USA}
\email{$^*$benlev@stanford.edu} 

\begin{abstract}
Digital Micromirror Devices (DMD) provide a robust platform with which to implement digital holography, in principle providing the means to rapidly generate propagating transverse electromagnetic fields with  arbitrary mode profiles at visible and IR wavelengths. We use a DMD to probe a Fabry-P\'{e}rot cavity in single-mode and near-degenerate confocal configurations. Pumping arbitrary modes of the cavity is possible with excellent specificity by virtue of the spatial overlap between the incident light field and the cavity mode.
\end{abstract}

\ocis{(090.1000) Aberration compensation; (090.1995) Digital holography;  (050.0050) Diffraction and gratings; (230.6120) Spatial light modulators; (140.3300) Laser beam shaping; (090.1760) Computer holography.}

\bibliographystyle{osajnl.bst}

\section{Introduction}

Cavity quantum electrodynamics (cQED) provides experimental access to the fundamental physics of light-matter interactions~\cite{walls2008quantum}.  Such systems are a useful tool in interferometry and spectroscopy\cite{Meiser:2009jg}, and quantum information storage and quantum simulation are promising  applications~\cite{Ritsch:2013fc}.
 
With few exceptions~\cite{Gangl:2000ff,Vuletic:2001bd,Horak:2002dw,Maunz:2003fu,Black:2003bd,Chan:2003bc}, two-mirror Fabry-P\'{e}rot cavities used in cQED research are probed via their fundamental TEM$_{0,0}$ mode.  However, recent interest in multimode cQED~\cite{Gopalakrishnan:2009gv,Gopalakrishnan:2010ey,Gopalakrishnan:2011jx,Strack:2011hv,Gopalakrishnan:2012cf,Muller:2012ju,Buchhold:2013fr,Torre:2013gc,Kollar2015,Schine:2015uj,Sommer:2015vn,Ningyuan:2015uj}---wherein the intracavity thermal or Bose-condensed atoms are simultaneously coupled to many degenerate but incommensurate optical modes---has encouraged an improved understanding of how cavity mode structure deviates from the ideal paraxial resonator as described in Ref.~\cite{siegman:86}.  In such experiments, the overlap between the atomic cloud and the cavity modes is a critical parameter that is dependent on the exact form of the cavity mode functions, the spectrum of which highly depends on mirror fabrication and alignment. 
 These overlaps are important to characterize, as they determine, e.g., the effective interaction length scales and atom-atom coupling strengths.   
 
We present here a method of mode-spectrum characterization based on driving the degenerate cavity with carefully tailored mode functions.  This enables the future  engineering of atom-atom interactions via shaping the pumped light fields and the measurement of the quantum system's dynamic response~\cite{Landig:2015et} to tailored drive fields.  Specifically, we show in this paper how a digital micromirror device (DMD), functioning as a binary spatial light modulator (SLM), can be used to generate holographic light fields that selectively couple to each cavity mode.  Such a device has recently been used to couple light into the modes of a ring resonator~\cite{Schine:2015uj}.  We show that these characterization tasks can be accomplished in a manner that is rapid and re-configurable using the DMD, enabling the  versatile \textit{in-situ} manipulation of one mode versus another. 

Furthermore, we study how the mode structure changes as a result of aberration mixing when the cavity is tuned to a degenerate configuration.  We demonstrate that the spatial properties of the intracavity field at multimode resonance can be controllably pumped.  Iterative procedures are employed to better couple to individual cavity modes; this paves the way for the use of optimization algorithms to automatically determine the cavity mode structure.

\section{Theory of the Paraxial Cavity}

The spectroscopic properties of an ideal Fabry-P\'{e}rot cavity are well described by paraxial wave theory~\cite{siegman:86}. The results of such analyses are two-fold: first, the modes themselves are Hermite-Gauss (HG) beams.  Indeed in practice, the fundamental mode of a cavity is nearly indistinguishable from an ideal Gaussian beam, even when using spherical, rather than parabolic, mirrors.  Second, the frequencies (in Hz) of these modes are given by
\begin{equation}
f_{qmn}= \frac{c}{2L}\left(q+ \frac{m+\frac{1}{2}}{\pi}\arccos\sqrt{g_{1x}g_{2x}}+
\frac{n+\frac{1}{2}}{\pi}\arccos\sqrt{g_{1y}g_{2y}}\right),
\end{equation}
where $c$ is the speed of light in the cavity, $L$ is the length of the two-mirror resonator, $g_{ij}=1-\frac{L}{R_{ij}}$, and $R_{ij}$ is the radius of curvature (ROC) of the $i^{th}$ mirror in the $j^{th}$ direction.  These denote orthogonal Cartesian axes. This formula captures many of the salient features of optical resonators. In a symmetric cavity where all ROCs are equal, the cavity may exhibit degenerate configurations at particular ratios of cavity length to ROC wherein a class of spatially incommensurate modes have identical resonant frequencies.   Specifically, this degeneracy can occur when the Gouy phase $\arccos(\pm g)$ is, e.g., 0 for planar cavities, $\pi/2$ for confocal cavities, and $\pi$ for concentric cavities.     

The ideal confocal cavity is one well-known example of a degenerate two-mirror resonator; a confocal cavity's length is equal to the ROC of its mirrors $L=R$. The mode-spectrum separates into two degenerate families alternating every half free spectral range (FSR), where an FSR is equal to $\frac{c}{2L}$. One degenerate family consists of all the even modes with $m+n\equiv0\mod 2$, while the other consists of the odd modes with $m+n \equiv 1 \mod 2$. It is worth noting that the azimuthal invariance in  ideal confocal cavities means that the eigenmodes are also well-described by Laguerre-Gauss functions, complex admixtures of the real-valued HG functions.\footnote{More generally, the Ince-Gauss modes provide a smooth crossover between these two families and are useful in describing real cavity modes where mode-mixing is present.}

The practicalities of mirror fabrication and mounting often break the mode degeneracy\footnote{Additionally, the two polarizations of each mode can be frequency-split by birefringence in the mirror and its coatings exacerbated by strain caused by the mounting supports, glue, etc.}; even if the mirrors are perfect paraboloids, they are unlikely to have the same four curvatures and same axial alignments. In these realistic cases, the spectroscopy of the near-confocal cavity is marked by two energy scales for the HG modes.  The first, $\Delta E_1=c(\delta_L-\delta_R)/\pi R^2$, is the frequency difference between families of different $m+n$, where $\delta_L=L-R$ and $\delta_R= \bar{R}_i - \bar{R}_j$.  Here, $\bar{R}_i = (R_{1i}+R_{2i})/2$, where $R_{1i}$ is the ROC of mirror 1 along direction $i$, etc. The second energy scale is $\Delta E_2 = \frac{c\delta_R}{2\pi R^2}$, which is the energy offset between modes within the same family. This picture is complicated by the fact that the cavity mirrors may not be mounted in such a way that the principle axes of the two mirrors are co-planar.  Moreover, mixing of the ideal HG modes into a non-uniform  $\Delta E_2$ spectrum  within modes of a family arises from aberrations in the shape of the mirror, e.g., spherical aberration, beyond-paraxial effects in the propagation of the light, and defects on the mirror surface.   This mixing increases as the modes begin to overlap in frequency as degeneracy is reached. \begin{figure}[t]
\centering\includegraphics[width=10cm]{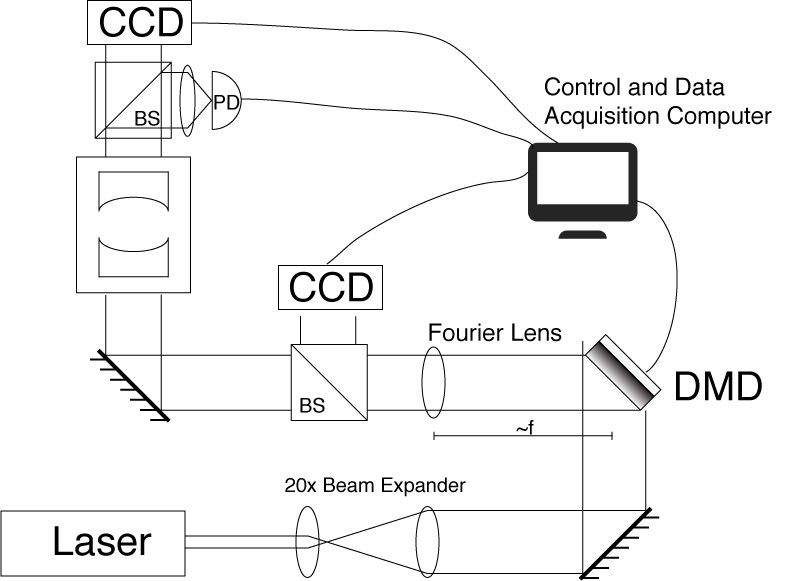}
\caption{Overview of the experiment. Laser light at~780.23~nm is incident on the DMD. The `on' pixels direct the light into the cavity via the first-order diffraction peak of the programmed holographic grating. The light reflected from the `off' pixels is caught by a beam block (not shown). The light emitted by the cavity is analyzed by a photodetector and a CCD camera.   The computer can analyse the images for mode content or generate the aberration-canceling phase mask. Software communicates a binary mask file to the DMD, called the DMD mask, for arbitrary field generation.}
\label{fig:OpticsDiagram}
\end{figure}

\begin{figure}[t]
\centering
\includegraphics[width=10cm]{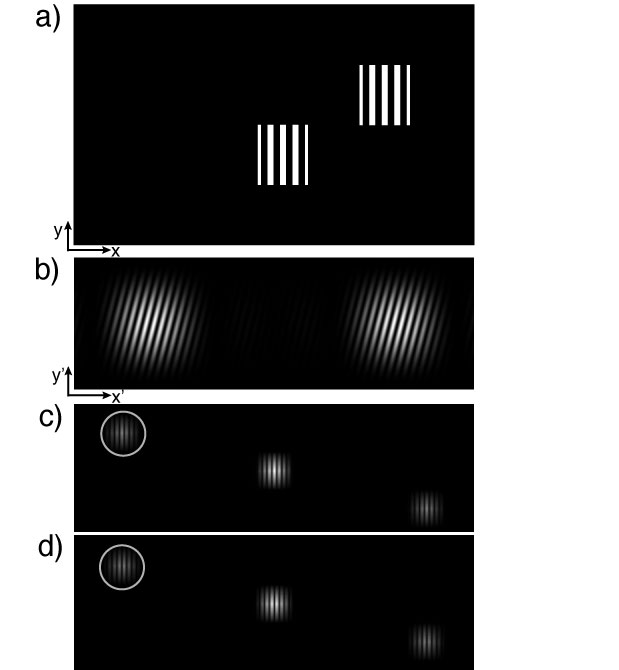}
\caption{An example of the window functions used in calibrating the DMD along with the associated interference patterns in the far field. a) Schematic of two activated regions containing a phase-coherent grating. In the actual DMD, the 912$\times$1140 pixels are grouped into 1,440 rectangular regions, each of which is 19$\times$38 pixels in extent. b) The interference pattern generated in the Fourier plane by a DMD programmed with the above mask. The phase of the fringe pattern reflects the  $\phi_{\Delta}(\Delta x,\Delta y)$ accumulated due to aberrations. c) The real and d) imaginary  parts of the numerical Fourier transform of the interference pattern shown in panel b).   The function describing the fringes in panel b) is $\cos^2(k_1 x' + k_2 y')$, meaning that in the Fourier space of panels c) and d), the spots are at $\pm(k_1,k_2)$ and $(0,0)$. The orientation of the fringes in each spot in c) and d) reflects the real space orientation of the DMD grating in panel a).  For ease, the phase $\phi_{\Delta}(\Delta x,\Delta y)$ is found not by analysing the fringes in panel b), but by taking the difference in phase between the pair of fringes marked by white circles in panel c) and d).}
\label{fig:Calibration_Figure}
\end{figure}

\section{DMD Calibration and Usage}

The DMD is positioned at one plane of a 2$f$ imaging system; in the paraxial limit, the electric field in this plane is connected to that in another plane by a Fourier transform. See Fig.~\ref{fig:OpticsDiagram}. This transform scales one set of spatial dimensions by $k/f$ with respect to the other, where $k=2\pi/\lambda$ is the wavenumber of the laser field. As such, a delta function in one plane $\delta(x-x_0)\delta(y-y_0)$ corresponds to a plane wave in the other plane $\exp[i k (x' x_0+y' y_0)/f]$, where $(x,y)$ and $(x',y')$ are pixel coordinates in the DMD and image planes, respectively.   It is this duality which allows us to use the DMD to generate arbitrary field profiles in the image plane using one of several algorithms known from digital holography~\cite{Lee:1974fw,Kuhn:14}. 

A necessary step before the DMD can be used to accurately produce field profiles  is the measurement of aberrations in the optical system, including the DMD itself \cite{Dholakia:10}.  The protective glass of the DMD is known to be under stress from manufacturing and thus exhibits non-uniform strain.  It is possible to characterize these aberrations by measuring the phase shift imparted onto a set of orthogonal basis functions generated at the DMD plane~\cite{Vellekoop:2007ko}.   Locally adding phase shifts in the DMD plane provides full phase control of the holographic image for either aberration removal, phase-sensitive image construction, or both simultaneously.  

For a binary amplitude mask like a DMD, one way to add phase to the field in the image plane is to include a grating into the mask design.  This is accomplished by using the DMD mask to modulate the amplitude of the field at the DMD plane via the grating function $\cos^{2}(k_x x + k_y y + \phi_m(x,y))$, where $\phi_m$ contains all the phase information of the mode and $k_x = kx_0/f$ and $k_y = ky_0/f$.  For simplicity, we choose approximate delta functions as our set of nearly-orthogonal basis functions since they are both orthogonal in the DMD plane and in any downstream plane. We approximate delta functions by partitioning the 912$\times$1140 mirror-pixels of the DMD into 1,440 individual 48$\times$30 rectangular regions.  Examples of two such windows are shown in Fig.~\ref{fig:Calibration_Figure}a.  The finite size of these window regions provides space within which to overlay phase-coherent grating functions.  The specific size is chosen as a balance between making them small enough to provide many partitions of the full DMD array---thereby providing high resolution for aberration correction---while also ensuring that each window is not too small, such that the transform-limited spot-size is not too large in the Fourier-image plane.  Setting $\phi_m(x,y)=0$ for the present discussion, each orthogonal basis function has the approximate form  of $\cos^{2}{\left(\frac{k}{f}(x x_0+y y_0)\right)}\Pi(\frac{x-x_c}{w})\Pi(\frac{y-y_c}{h})$ at the DMD plane, where $\Pi(x)$ is the rectangle function, $(x_c,y_c)$ is the center of the region, and $w$ and $h$ are the width and height  of the region, respectively. Turning on any two windows---defined by differing $(x_c,y_c)$---yields two interference patterns in the image plane centered at $\pm(2x_0,2y_0)$, as shown in Fig.~\ref{fig:Calibration_Figure}b.  Each of these has the form $\cos^{2}(\frac{k}{f} (x' \Delta x + y' \Delta y))$, where $\Delta x$ and $\Delta y$ are the separation in $x$ and $y$ in the DMD plane between the two centers of the window regions.

Aberrations in the imaging system are determined by measuring  the phase shift in the image plane acquired by light reflected from these window regions in the DMD plane. In the presence of aberrations in the optical path, the interference pattern shown in Fig.~\ref{fig:Calibration_Figure}b is modified to be $\cos^{2}(\frac{k}{f} (x' \Delta x + y' \Delta y)+\phi_{\Delta} (\Delta x,\Delta y))$, where the phase $\phi_{\Delta}(\Delta x,\Delta y)$ is that which is accumulated due to the particular optical aberration encountered by the light propagating from the window region to the image plane.  The phase of this pattern is easily obtained by taking the Fourier transform of the interference pattern, which produces three complex-valued intensity features.  These can be split into a real-valued image and an imaginary-valued image, as shown in Figs.~\ref{fig:Calibration_Figure}c and d. The phase $\phi_{\Delta}(\Delta x,\Delta y)$ is found by taking the difference in phase between the pair of fringes marked by white circles in Figs.~\ref{fig:Calibration_Figure}c and d.  The $\phi_{\Delta}(\Delta x,\Delta y)$ is then measured for all windows tiling the DMD pixel area.  We employ this procedure by fixing one window in the center of the DMD pixel array while another window is sequentially tiled over all other window regions.   Together this allows one to determine a phase mask that is a full representation of the linear aberrations of the optical system.  The inverse of such a mask can be superimposed onto any hologram mask programmed into the DMD for the purpose of cancelling these aberrations.  In this fashion we are able to produce high-fidelity representations of desired field mode profiles.

\section{Experiment}
\subsection{Experimental outline}
The DMD used is a Texas Instruments DLP4500NIR consisting of 912$\times$1140 individually addressable square aluminum mirrors, which measure $\sim$10.8~$\mu$m on the diagonal.  They are arranged on a square lattice and can be independently controlled by sending an array of 1's (`on') and 0's (`off') to the controller.  Physically, these values correspond to the mirrors changing their angle by $\pm 12^\circ$ with respect to the substrate on which they are fabricated.  Light reflected from `off' mirrors is directed into a beam-blocker, while light reflected from the `on' mirrors is used to form the desired hologram and cavity in-coupling light. We generate a set of DMD masks for each experiment we wish to perform; the rate at which we can update holograms is presently limited by the 60-Hz refresh rate of the software driver. The laser beam incident on the DMD surface is magnified to a $1/e$ radius of 8~mm, quite a bit larger than the 7-mm extent of the 45$^\circ$-projected DMD surface.  This insures that the DMD is exposed to an homogeneous light intensity across its surface, which we verify by imaging the surface.

The 2-cm cavity mirrors are glued onto a rotationally symmetric brass mount for temperature and vibration stability. The cavity length can be adjusted via the turning of an 80-turns-per-inch screw that holds one of the two mirrors. This screw provides course control of the cavity length: $4.4 \ \mu \mathrm{m}/5 ^\circ$. Glued in between the screw and the mirror is a single-crystal PZT that can be voltage-tuned through $210 \ \mathrm{nm}$. The lens connecting the two Fourier planes has a focal length of $f=200 \ \mathrm{mm}$, chosen to maximize the DMD area used in producing the HG fields that will be mode-matched to the cavity. Figure~\ref{fig:OpticsDiagram} presents  a diagram of the setup used to couple holograms generated by the DMD into the cavity. The finesse of the cavity is $\sim$$5\times10^4$ at the wavelength we use for this work, 780 nm, resulting in a cavity linewidth of $\sim$300~kHz.

\begin{figure}[t]
\centering
\includegraphics[width=14cm]{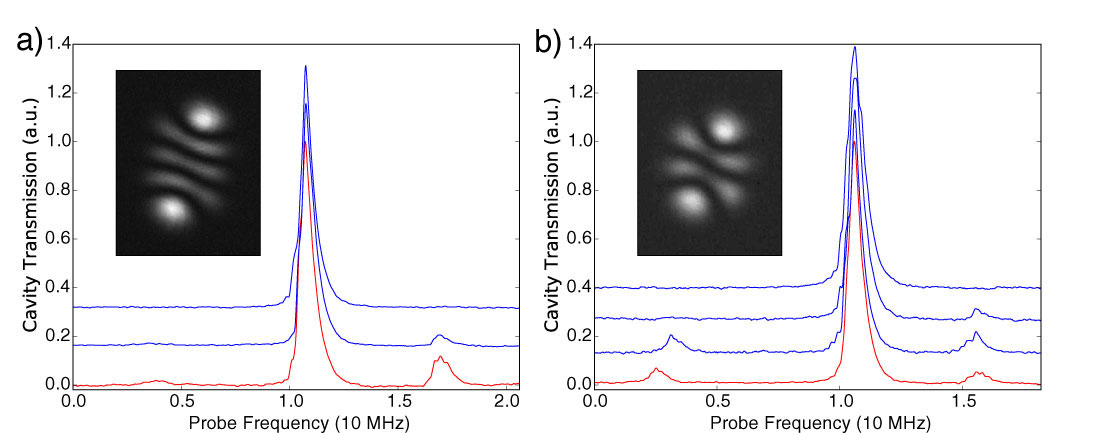}
\caption{Cavity transmission versus pump frequency.  a) Coupling to the cavity (3,1) mode. Native mixing in the cavity results in mixing of the ideal cavity modes such that the (3,1) does not equal the ideal HG$_{3,1}$. The red transmission curve results from in-coupling an ideal HG$_{3,1}$ mode into the cavity. The imperfect overlap induces a coupling to the adjacent modes in addition to the targeted  (3,1). The blue curves show the result of increasing optimization of the DMD mask, resulting in a hologram that couples exclusively to the cavity (3,1) mode, which is shown in the inset.  We determine that the cavity (3,1) mode roughly consists of  8\% HG$_{2,2}$, 8\% HG$_{4,0}$, and 84\% HG$_{3,1}$. b) Same is in panel a), but with respect to the (2,1) mode.  We determine the cavity (2,1) mode roughly  consists of   4\% HG$_{1,2}$, 4\% HG$_{3,0}$, and 92\% HG$_{2,1}$. }
\label{fig:Rectification}
\end{figure}

\begin{figure}[t]
\centering
\includegraphics[width=14cm]{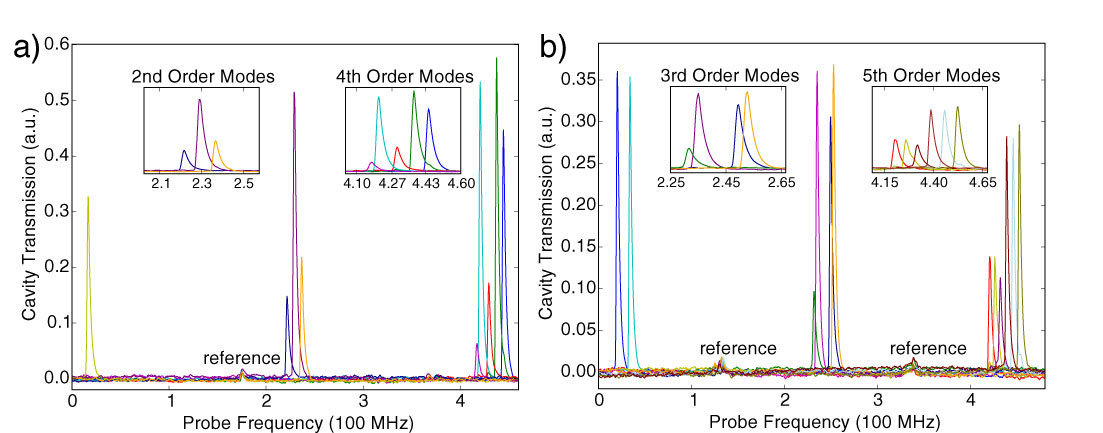}
\caption{Cavity transmission versus pump frequency for different  DMD  masks.  Each trace corresponds to a distinct mask generating a unique and orthogonal hologram on the DMD. a) Even mode families.  Insets show the cavity transmission for the 2nd and 4th-order families. The feature above the noise floor marked `reference' was created with a superimposed Gaussian beam in order to align the transmission curves to one another. b) Same as in panel a), but odd mode families. Insets show the cavity transmission for the 3rd and 5th-order families of modes.}
\label{fig:single}
\end{figure}

\subsection{Designing a DMD mask for a particular HG mode}

The HG modes are a good basis set for DMD studies because they are eigenfunctions of the Fourier transform.  That is, if the field created by the mask at the DMD plane has the form of an HG mode, then the field in the associated Fourier image plane will be a scaled version of the same HG mode ~\cite{Zupancic:2003}.

To couple light to the cavity, it is necessary to create HG modes with the appropriate scale using the DMD.
We first choose a target HG mode $\Phi_{HG}(x,y)$ normalized such that $\max{|\Phi_{HG}(x,y)|} = 1$.  The phase of the mode $\phi_{HG}(x,y)$ is superimposed with the aberration-correcting phase such that the phase of the hologram to be programmed into the DMD is $\phi_m(x,y)=\frac{1}{2}(\phi_{HG}(x,y) - \phi_\Delta(x,y))$. Both phases are normalized to run between 0 and $2\pi$. 
Including the grating structure necessary to create the HG mode in the Fourier image plane, the final field to be produced by the DMD mask is
\begin{equation}\label{field_dmd}
E_{DMD} = |\Phi_{HG}(x,y)|\cos\left(\frac{k}{f}[x x_0+y y_0+\frac{z_0}{2f} C(x,y)] + \phi_m(x,y)\right)^2,
\end{equation}
where $kx_0/f$ and $ky_0/f$ determine the periodicity of the grating in the DMD plane and thus the center location of the mode in the image plane at $\pm(2x_0,2y_0)$. $z_0$ is a focusing parameter and $C(x,y)$ is a focusing field that is in the form of a paraboloid that corrects for the 45$^\circ$-angle of the DMD face.  
The function in Eqn.~\ref{field_dmd} is sampled using Floyd-Steinberg dithering with a two-color palette to numerically develop a pixel-by-pixel representation of this field at the DMD surface for programming into the controller.  A final reshaping is necessary to generate the file that correctly maps pixel values to DMD mirrors. This file is then loaded into the DMD controller.  The whole process takes only 30~s per mode-image using a commercial laptop running code written in Python.

\subsection{Mode-coupling procedure}

To couple a mode into the cavity, we first choose $x_0$ and $y_0$ to be large enough to sufficiently separate the grating orders in the Fourier image plane.  The first-order peak contains the mode image to be coupled and is  directed into the cavity while the other orders are blocked.  Transmission through the cavity is maximized using two mirrors placed immediately before the cavity. The Gaussian waist of the incoming mode and $z_0$ are optimized through manual iteration to maximize the coupling of the mode.   For simplicity, we typically use the fundamental HG$_{0,0}$ 
mode for this procedure, which fixes all but one of the cavity parameters. The indices are the number of nodes in the $i$ and $j$ axes, respectively.  

The remaining parameter, the azimuthal angle of the nodal axes in the transverse plane, is determined manually by observing which modes in a set of non-azimuthally symmetric modes maximize coupling.  For simplicity, we use a collection of HG$_{1,0}$ beams at different azimuthal angles $\theta$, finally choosing the $\theta$ corresponding to the HG$_{1,0}$ mode exhibiting the best coupling efficiency. 

This procedure would, in principle, fully determine the HG mode-basis of the cavity if the cavity conformed to the ideal paraxial model.  However, intrafamily mixing is  present, as mentioned above, and the actual cavity modes do not always correspond to ideal HG-mode shapes.  An example of this can be seen in the insets of Fig.~\ref{fig:Rectification}. To label the modes comprising the actual mode spectrum, we number each mode of an $n+m=N$ family by indices $(\alpha,\beta)$, where $\alpha$ and $\beta$ no longer necessarily correspond to the number of nodes along a direction $i$ and $j$, respectfully.  However, the condition $\alpha + \beta = N$ remains fixed for each mode family $N$.\footnote{We note that the 210-MHz splitting between families ensures that inter-family mixing is negligible, thereby ensuring that $\alpha + \beta = n+m = N$ is a good approximation.}   Moreover, we choose to denote the highest-frequency mode in a family the $(0,N)$ mode, rather than the $(N,0)$ mode.\footnote{Among the mode families we explore here, the modes with $N$ nodes along only one direction typically possess the highest resonance frequency.}

\begin{figure}[t]
\centering
\includegraphics[width=12cm]{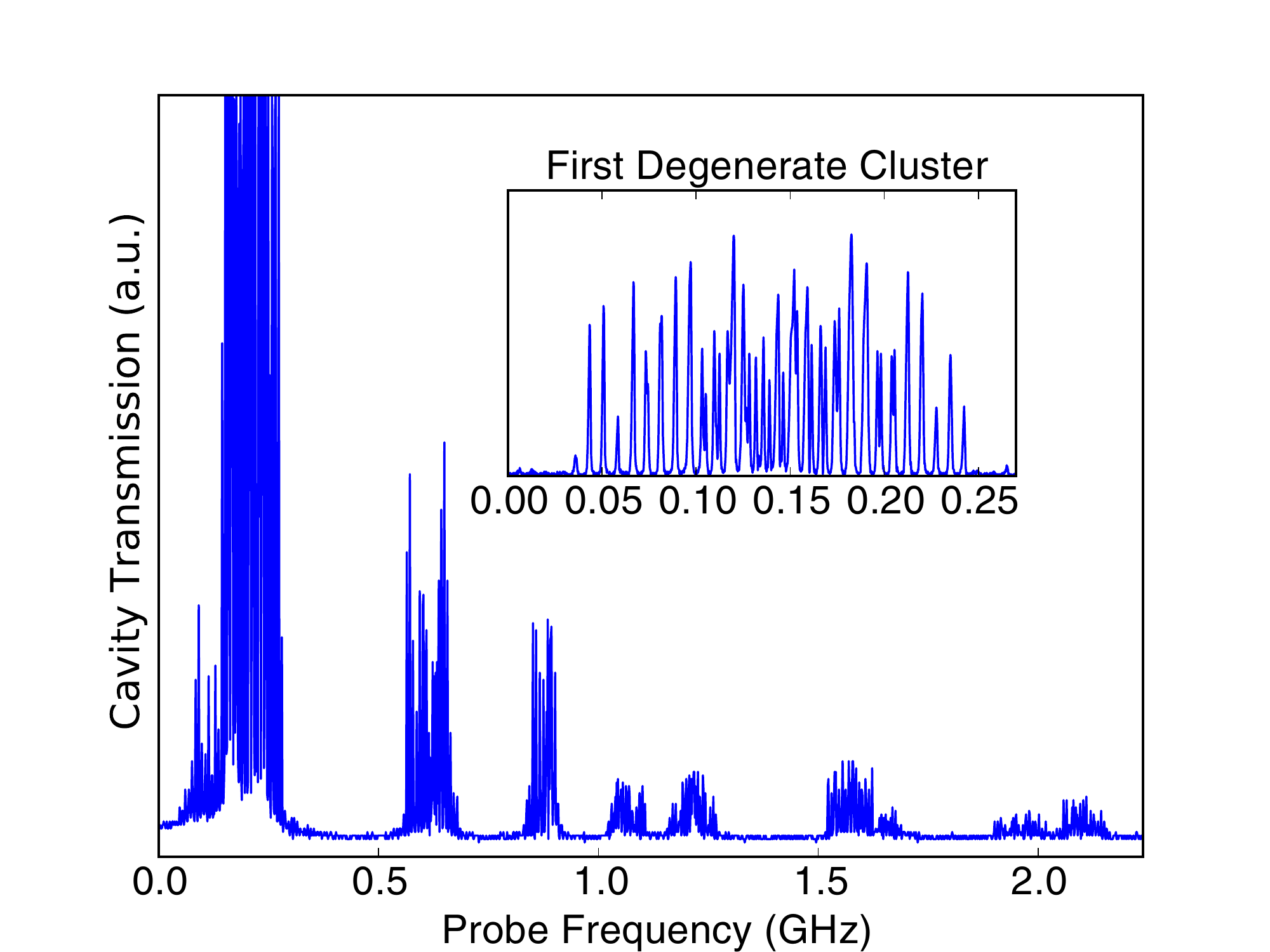}
\caption{Spectroscopy of the cavity in a near-degenerate configuration wherein the first even mode becomes degenerate with the fundamental.  Inset: zoomed-in spectrum of the first clump of degenerate modes.}
\label{fig:all-modes}
\end{figure}

\begin{figure}[t]
\centering
\includegraphics[width=14cm]{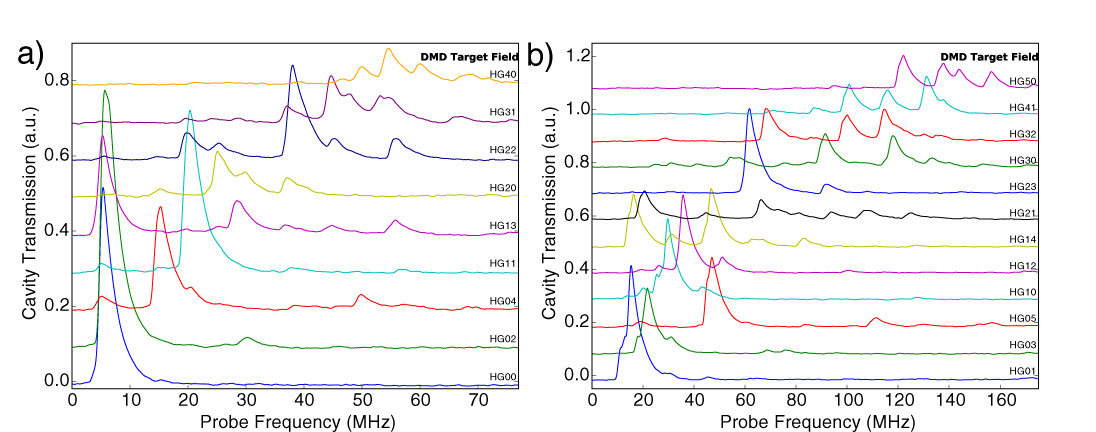}
\caption{Cavity transmission when the cavity is pumped by fields created by DMD masks of ideal HG-modes. a) Even mode spectra. b) Odd mode spectra.}
\label{fig:mm}
\end{figure}

\subsection{Single-mode spectroscopy}

The single-mode character of the cavity is probed by setting the length of the cavity to be 1-mm-longer than the confocality condition. 
While most modes in a given family possess a spatial structure very close to an ideal HG-mode, we do find at least one per family that significantly deviates from the HG-modal structure. For example, in the $N=3$ and $N=4$ families, the $(2,1)$ and $(3,1)$ modes both show significant deviation from the ideal HG$_{2,1}$ and HG$_{3,1}$ modes, respectively, as can be seen from the insets of Fig.~\ref{fig:Rectification}. We determine the HG-modal decompositions through a 2D fitting routing, and the fraction of other intrafamily HG modes is listed in the captions.

Figure~\ref{fig:Rectification} shows the results of an iterative procedure used to probe the $(2,1)$ and $(3,1)$ modes. In each case, attempting to couple to the mode with an ideal HG resulted in cross-coupling to the immediately adjacent modes. The optimized masks consist primarily of the HG mode in question with small (5\%-10\%) admixtures of the adjacent HG modes. Each mask was rotated to an angle $\theta$ optimal for mode coupling.   Though there can be much mixing of HG modes, we are able to successfully generate holograms that couple to the $(2,1)$ and $(3,1)$ cavity modes with high specificity.  As seen in Fig.~\ref{fig:Rectification}, all transmission in  non-targeted modes are below the level of our detection signal-to-noise. Although optimizing by hand is quite straightforward, an automated procedure can be applied to determine the eigenmodes of the cavity more quickly.

Other than the (2,1) and (3,1) modes, all other modes in these families were efficiently coupled to by DMD masks employing the ideal HG mode functions. The result of a spectroscopic scan of the low-order odd and even mode families is shown in Fig.~\ref{fig:single}. Each trace corresponds to a separate DMD mask designed to couple to a specific single mode of the cavity. Our manual optimization routine ensures that unwanted cross-coupling is minimal and indeed not apparent in these traces.  All modes emitted out of the cavity nearly identical to ideal HG modes, with the exception of the two aforementioned mixtures. We note that the rapidity of the frequency sweep results in an asymmetric lineshape for each mode due to ringdown effects.
\begin{figure}[t]
\centering
\includegraphics[width=9cm]{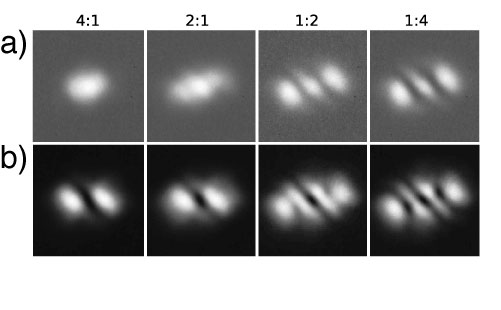}
\caption{a) Four transmission images of the near-degenerate cavity when pumped at a particular  frequency by a mask consisting of an increasing amount of an HG$_{2,0}$ mode added to an HG$_{0,0}$ mode.  Each image is the light emitted from the cavity and is the coherent addition of the two modes at a single probe frequency.  b) Same as panel a), but with an increasing amount of an HG$_{3,0}$ mode added to an HG$_{1,0}$ mode.}
\label{fig:mix}
\end{figure}

\subsection{Multimode Spectroscopy}

We experimentally probe the basic properties of the near-degenerate cavity at hand, because a numerical characterization~\cite{Vinet:2007gx} is impractical due to the need to minutely characterize the mode mixing arising from mirror defects, etc.
To proceed, the cavity is tuned to a near-degenerate configuration wherein the length was adjusted until the nearest second-order mode (0,2) became degenerate with the fundamental (0,0).\footnote{Due to misalignment of the cavity mounts, this cavity length was the point at which the maximum mode density what achieved. Other cavities can achieve far greater mode density, e.g., see Ref.~\cite{Kollar2015}.}  Figure \ref{fig:all-modes} illustrates the character of the degeneracy in this configuration.  To allow the observation of the frequency spacing of modes, the cavity was pumped with a broad beam slightly off-center.  The modes are seen to arrange into clusters in this near-confocal configuration.  It is possible to study some aspects of cavity degeneracy in this configuration since many modes appear with separations of less than a few linewidths.

The transmission of the cavity was measured for each ideal HG mask as before. Figure~\ref{fig:mm} shows the cavity transmission for each mask. As expected, there is a now a greater degree of mode mixing, especially in the higher order modes, as can be seen by the prevalence of side-peaks at many distinct probe frequencies. Nevertheless, each ideal HG mask couples to a small collection of cavity modes within its spectral vicinity.  We note that the two lowest-order even and odd modes---blue traces at the bottom of each figure---can be selectively addressed by distinct, orthogonal HG masks. 
Finally, to demonstrate that the near-degenerate cavity can be pumped by tailored field modes, we show in Fig.~\ref{fig:mix} that any linear combination of, e.g., HG$_{0,0}$ with HG$_{2,0}$ or HG$_{1,0}$ with HG$_{3,0}$, can be coupled into the near-degenerate cavity.

\section{Conclusion}

The ability of cavity fields to mediate  atomic interactions with high flexibility is an enticing new direction in quantum simulation.  The DMD has allowed us to demonstrate an important step in this direction, namely the identification and specific probing of modal content within overlapping mode spectra of single-mode and near-degenerate confocal Fabry-Per{\'{o}}t cavities.

\section*{Acknowledgments}

The authors thank J.~Keeling and T.~Ciobanu for helpful discussions.  Funding for this effort was provided by the Army Research Office (ARO), and B.L.L.~also thanks the David and Lucile Packard Foundation for support.

\end{document}